\batchmode
\documentclass[10pt,a4paper,english,floatfix,aps,prb,superscriptaddress,reprint]{revtex4-1}
\usepackage[T1]{fontenc}
\usepackage[latin9]{inputenc}
\usepackage{amsmath}
\usepackage{amssymb}
\usepackage{graphicx}
\usepackage{esint}

\makeatletter

\pdfpageheight\paperheight
\pdfpagewidth\paperwidth

\@ifundefined{textcolor}{}
{%
 \definecolor{BLACK}{gray}{0}
 \definecolor{WHITE}{gray}{1}
 \definecolor{RED}{rgb}{1,0,0}
 \definecolor{GREEN}{rgb}{0,1,0}
 \definecolor{BLUE}{rgb}{0,0,1}
 \definecolor{CYAN}{cmyk}{1,0,0,0}
 \definecolor{MAGENTA}{cmyk}{0,1,0,0}
 \definecolor{YELLOW}{cmyk}{0,0,1,0}
}

\usepackage{subfigure}

\makeatother

\usepackage{babel}
\begin{document}

\title{Perturbative approach to the capacitive interaction between a  sensor
quantum dot and a charge qubit}

\author{S. Mojtaba Tabatabaei}

\email{sm_tabatabaei@sbu.ac.ir}

\selectlanguage{english}%

\affiliation{Department of Physics, Shahid Beheshti University, Tehran, Iran}
\begin{abstract}
We consider the capacitive interaction between a charge qubit and
a sensor quantum dot(SQD) perturbatively to the second order of their
coupling constant at zero temperature by utilizing the method of non-equilibrium
Green's functions together with infinite-U Lacroix approximation and
employing Majorana fermion representation for qubit isospin operators.
The effect of back-actions on dynamics of the system is taken into
account by calculating the self-energies and the Green's functions
in a self-consistent manner. To demonstrate the applicability of the
method, we investigate relevant physical quantities of the system
at zero and finite bias voltages. In the regime of weak SQD-qubit
coupling, we find a linear relation between the stationary-state expectation
values of the third component of the qubit isospin vector, $\left\langle \tau_{3}\right\rangle $,
and the differential conductance of the SQD. Furthermore, our numerical
results predict that the effect of SQD-qubit coupling on differential
conductance of the SQD should be maximized at zero bias voltage. Moreover,
we obtain an analytical expression to describe the behavior of the
differential conductance of the SQD with respect to the qubit parameters.
Our results at zero bias voltage are consistent with the results of
numerical renormalization group method.
\end{abstract}


\maketitle

\section{Introduction}

Typically, the state of a solid state qubit could be indirectly extracted
by measuring the conductance of a current carrying electro-meter which
is capacitively coupled to the qubit. This detector, which could be
realized in the experiment by a quantum point contact(QPC)\cite{aleiner1997,gurvitz1997,nature02693,korotkov2001continuous,goan2001continuous,stace2004continuous,engel2004measurement,luo2013}
or a single electron transistor(SET)\cite{schoelkopf1998,nakamura1999coherent,aassime2001radio,lu2003,astafiev2004,lahaye2004approaching,turek2005single,barthel2010},
provides us with measurements of the charge fluctuations of qubit.
The usage of SETs are however more advantageous to the QPCs because
of their much more sensitivity to the charge fluctuations\cite{barthel2010}.
In practice, the coupling of the sensor quantum dot(SQD) of the SET
with qubit is made so weak in order to reduce the effect of measurements
on the qubit state. But, no matter how much it is weak, the system
inevitably suffers from coupling effects which results in a coherent
back-action on the qubit dynamics and renormalization of the system
energy levels.\cite{zorin1996,grishin2005}

Using SET as a qubit detector has been the subject of several theoretical
studies\cite{makhlin2001quantum,PhysRevLett.84.5820,makhlin2000statistics,korotkov2001selective,gurvitz2005,gurvitz2008,shnirman1998,mozyrsky2004quantum,oxtoby2006sensitivity,emary2008quantum,schulenborg2014detection,hell2014,hell2016}.
Much works have been devoted to investigate the time dependent dynamics
of the reduced density matrix of the system with considering the leading
order tunneling processes in the SET and ignoring the back-action
of qubit and SET on each other\cite{makhlin2000statistics,korotkov2001selective,gurvitz2005,gurvitz2008}.
The problem of considering the effects of back-actions on the SET-qubit
system in the presence of external bias were also studied in Refs.\onlinecite{shnirman1998,mozyrsky2004quantum,oxtoby2006sensitivity,emary2008quantum,schulenborg2014detection,hell2014,hell2016}.
Recently, Hell et al.\cite{hell2014,hell2016} studied the coherent
back-action of the measurements on the SET-qubit system in the presence
of finite bias by deriving Markovian kinetic equations for the system
with considering next to the leading order corrections into the tunneling
processes of the SET and the effects of energy levels's renormalization
of the system.

Here, we consider the application of the method of non-equilibrium
Green's functions for describing the non-equilibrium dynamics of the
SET-qubit system. We calculate the steady-state non-equilibrium Green's
functions of the system at zero temperature using second order self-energies
of the capacitive coupling between SQD and qubit. Due to the lack
of applicability of the Wick's theorem for Pauli operators, we utilize
the Majorana Fermion representation\cite{tsvelik1992new,mao2003spin,shnirman2003spin}
for the qubit isospin operators by which a systematic diagrammatic
perturbative expansion of the system's Green's functions become possible.
In order to take into account strong electron-electron interaction
on the SQD, which is necessary to keep it in the Coulomb blockade
regime, we employ the infinite-U Lacroix approximation\cite{lacroix1981density}
to calculate the bare Green's functions of SQD. The back-action effects
on the average occupations of SQD and qubit are accounted for by calculating
the self-energies and the Green's functions self-consistently. Using
the calculated interacting Green's functions of the system, we investigate
the density of states of the SQD and the steady-state expectation
value of the third component of the isospin operator of the qubit,
$\left\langle \tau_{3}\right\rangle $. Furthermore, we determine
the differential conductance of SQD at zero and finite bias voltages
and show that there is a linear relation between SQD's differential
conductance and the steady-state expectation value $\left\langle \tau_{3}\right\rangle $.
We check the accuracy of our results at zero bias by comparing them
with the results obtained from numerical renormalization group(NRG)
method\cite{bulla2008numerical}.

Our approach differs basically from density matrix based approaches\cite{breuer2002theory}.
In the latter, it is the coupling of SQD with electrodes which is
considered perturbatively for calculating the reduced density matrix
of the subsystem and the possible partial coherencies between different
charge states of the SQD during tunneling processes are ignored trivially.
Instead, in our approach, the parameter that is used for perturbatively
obtaining the Green's functions of the system is the capacitive coupling
between SQD and qubit while the effects of coupling between SQD and
metallic electrodes are incorporated non-perturbatively into the Green's
functions of SQD, which retains the possible partial coherencies of
SQD's charge states. 

The paper is organized as follows: In Sec. \ref{sub:Model-Hamiltonian},
the model Hamiltonian is presented. Then in Sec. \ref{sub:Non-equilibrium-Green's},
we present the derivation of non-equilibrium Green's functions of
the SQD-qubit system. After that, in Sec.\ref{sub:Physical-quantities},
we give some expressions for relating different physical quantities
of the system with the Green's functions. We present our numerical
results in Sec. \ref{sec:Numerical-results}. Then, in Sec. \ref{sec:conclusions},
we give a summary of our work and some concluding remarks related
to it.

\section{Theoretical formalism}

\subsection{\label{sub:Model-Hamiltonian}Model Hamiltonian}

\begin{figure}
\includegraphics[width=8.6cm]{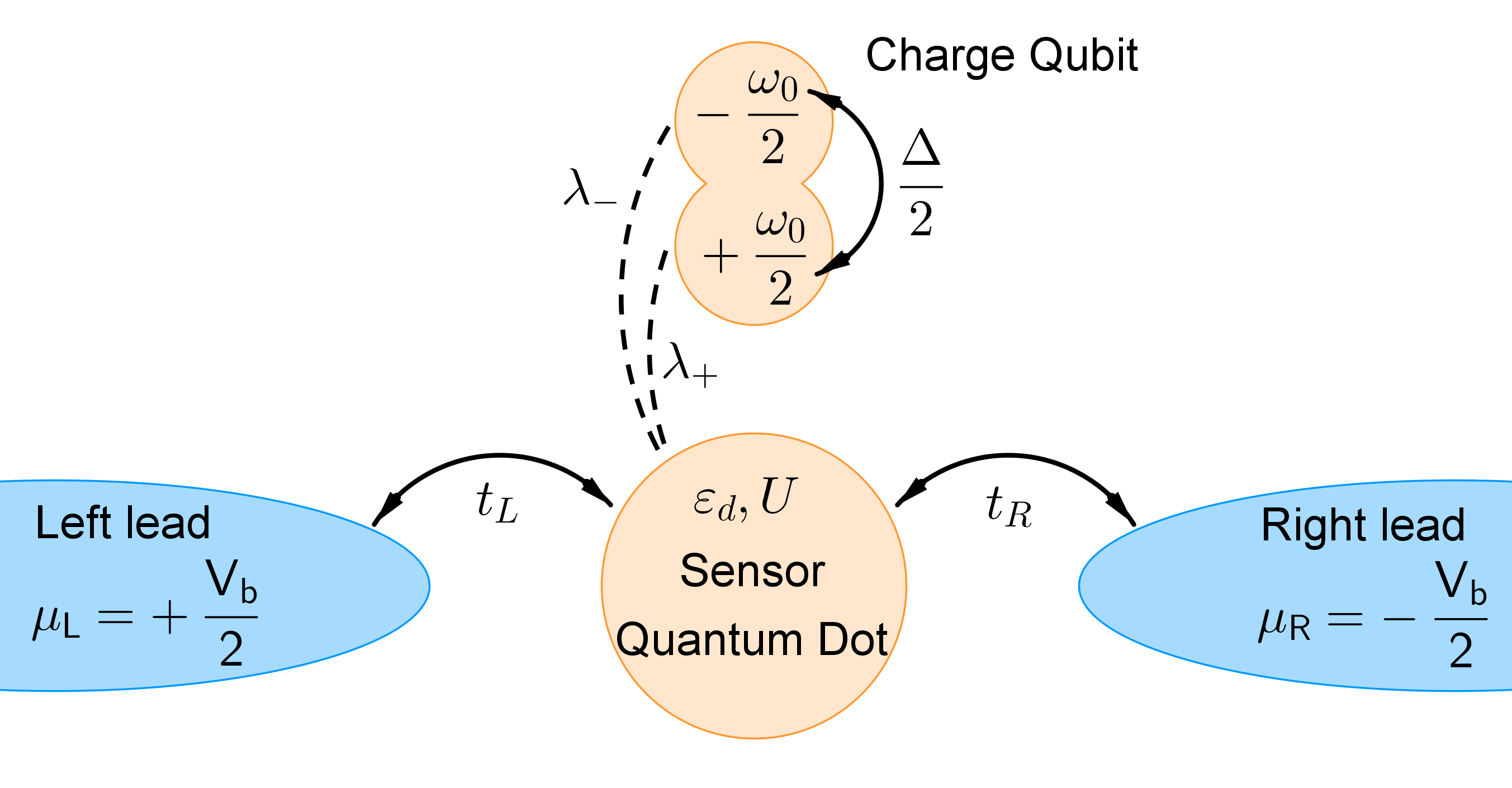}

\protect\caption{\label{figsys}(Color online) Schematic representation of the model
system. The sensor quantum dot, characterized with gate voltage $\varepsilon_{d}$
and on-site interaction $U$, coupled to two metallic leads and simultaneously
interacts capacitively with one double quantum dot through interaction
constants $\lambda_{+}$ and $\lambda_{-}$. The on-site energies
of DQD are $\pm\omega_{0}$ and the tunneling energy between its dots
is specified by $\Delta/2$.}
\end{figure}
Our model system, as is depicted in Fig.\ref{figsys}, consists of
a SQD in a Coulomb blockade regime tunnel coupled to two metallic
electrodes while simultaneously interacts capacitively with a charge
qubit which is modeled by a double quantum dot(DQD). The total Hamiltonian
of the system can be written as 
\begin{equation}
\mathcal{H}=\mathcal{H}_{SET}+\mathcal{H}_{DQD}+\mathcal{H}_{I}.\label{H_total}
\end{equation}

The first term is the Hamiltonian of SET which is given by 
\begin{alignat}{1}
\mathcal{H}_{SET} & =\underset{\nu}{\sum}\varepsilon_{d}\, c_{d,\nu}^{\dagger}c_{d,\nu}+Un_{d,\uparrow}n_{d,\downarrow}+\nonumber \\
 & \underset{k,\alpha,\nu}{\sum}(\varepsilon_{k}+\mu_{\alpha})\, c_{k,\alpha,\nu}^{\dagger}c_{k,\alpha,\nu}+t_{\alpha}\,\left(c_{k,\alpha,\nu}^{\dagger}c_{d,\nu}+H.c\right),\label{eq:H_set}
\end{alignat}
where the operator $c_{d,\nu}^{\dagger}\left(c_{d,\nu}\right)$ creates(annihilates)
an electron with spin $\nu=\uparrow,\downarrow$ in the SQD, $n_{d,\nu}=c_{d,\nu}^{\dagger}c_{d,\nu}$
is the spin dependent electron occupation operator of SQD, $\varepsilon_{d}$
is the applied gate voltage and $U$ is the on-site electron-electron
interaction energy in SQD. Similarly, the operator $c_{k,\alpha,\nu}^{\dagger}$$\left(c_{k,\alpha,\nu}\right)$
is the corresponding operator for electron creation(annihilation)
with energy $\varepsilon_{k}$ in the left and right leads ($\alpha=L,R$),
each of which are treated as half-filled quasi-one-dimensional normal
metals with chemical potentials $\mu_{L}$ and $\mu_{R}$, respectively.
The coupling of the SQD with each lead is assumed to be energy and
spin independent and characterized by a hybridization constant, $t_{L,R}$.

The second term in Eq.(\ref{H_total}) is the Hamiltonian of charge
qubit, which is modeled as a double quantum dot\cite{gorman2005charge}
containing only one electron, with on-site energies $\pm\frac{\omega_{0}}{2}$
and hybridization energy $\frac{\Delta}{2}$. Representing the state
of the electron on each of DQD's dots by $\left|+\right\rangle $
and $\left|-\right\rangle $, in terms of isospin operators $\left(\tau_{1},\tau_{2},\tau_{3}\right)$
the DQD's Hamiltonian takes the form 
\begin{align}
\mathcal{H}_{DQD} & =-\frac{\omega_{0}}{2}\tau_{3}+\frac{\Delta}{2}\tau_{1}.\label{eq:H_dqd}
\end{align}

The last term in Eq.(\ref{H_total}) is the capacitive interaction
between SQD and DQD which is $2n_{d}(\lambda_{+}n_{+}+\lambda_{-}n_{-})$,
where $\lambda_{+,-}$ is an interaction constant and $n_{d}=n_{d,\uparrow}+n_{d,\downarrow}$
is the total electron number operator of SQD. Furthermore, $n_{+,-}$
represents the occupation number operator of each dots of the DQD.
By using the relation $n_{\pm}=\frac{1}{2}(1\pm\tau_{3})$ and an
appropriate renormalization of $\varepsilon_{d}$, the interaction
can be expressed as $\lambda n_{d}\tau_{3},$ where $\lambda\equiv\lambda_{+}-\lambda_{-}$.
In addition, for later convenience, we explicitly take into account
the mean-field back-action effects by adding and subtracting the operator
$A=\lambda\left(\left\langle n_{d}\right\rangle \tau_{3}+\left\langle \tau_{3}\right\rangle n_{d}\right)$
to the total Hamiltonian. This modifies the on-site energies of the
SQD and DQD to $\tilde{\varepsilon}_{d}=\varepsilon_{d}+\lambda\left\langle \tau_{3}\right\rangle $
and $\tilde{\omega}_{0}=\omega_{0}-2\lambda\left\langle n_{d}\right\rangle $,
respectively, and then the interaction term of Hamiltonian becomes
\begin{gather}
\mathcal{H}_{I}=\lambda(n_{d}-\left\langle n_{d}\right\rangle )(\tau_{3}-\left\langle \tau_{3}\right\rangle ).\label{eq:H_qd-dqd}
\end{gather}

In order to use perturbation theory we need to express the isospin
operators in terms of Majorana fermion operators by\cite{shnirman2003spin}
\begin{equation}
\tau_{a}=-i\epsilon_{abc}\eta_{b}\eta_{c},\label{eq:sigma2majorana}
\end{equation}
for $a,b,c=1,2,3$, where $\epsilon_{abc}$ is the Levi-Civita antisymmetric
tensor and $(\eta_{1},\eta_{2},\eta_{3})$ are three Majorana fermion
operators satisfying usual fermionic equal-time anti-commutation relation
$\{\eta_{a},\eta_{b}^{\dagger}\}=\delta_{a,b}$ with $\eta_{a}^{\dagger}=\eta_{a}$.

\subsection{\label{sub:Non-equilibrium-Green's}Non equilibrium Green's functions}

The non-equilibrium Green's function method is a usual choice to study
out of equilibrium systems\cite{stefanucci2013nonequilibrium}. We
will treat $\mathcal{H}_{SQD}+\mathcal{H}_{DQD}$ and $\mathcal{H}_{I}$
as the non-interacting and interaction parts of Hamiltonians, respectively.
A complete description of the non-equilibrium steady-state of a system
requires the knowledge of four Green's functions, which we choose
the retarded, advanced, lesser and greater, defined, respectively,
for the non-interacting system as\begin{subequations}\label{GF} 
\begin{gather}
g_{s,mn}^{R}\left(t,t'\right)=-i\theta(t-t')\left\langle \left\{ \Psi_{s,m}\left(t\right),\Psi_{s,n}^{\dagger}\left(t'\right)\right\} \right\rangle _{0},\\
g_{s,mn}^{A}\left(t,t'\right)=i\theta(t'-t)\left\langle \left\{ \Psi_{s,m}\left(t\right),\Psi_{s,n}^{\dagger}\left(t'\right)\right\} \right\rangle _{0},\\
g_{s,mn}^{<}\left(t,t'\right)=i\left\langle \Psi_{s,n}^{\dagger}\left(t'\right)\Psi_{s,m}\left(t\right)\right\rangle _{0}\label{eq:G_les}
\end{gather}
and
\begin{equation}
g_{s,mn}^{>}\left(t,t'\right)=-i\left\langle \Psi_{s,m}\left(t\right)\Psi_{s,n}^{\dagger}\left(t'\right)\right\rangle _{0},
\end{equation}
\end{subequations}where $s=d,\eta$ determines the corresponding
subsystem for which the Green's functions are defined and $m,n$ represent
degrees of freedom for the corresponding subsystem, that is, in the
case of SQD, $\Psi_{d,m}=c_{d,m}$ with $m=\uparrow,\downarrow$ while
for DQD, $\Psi_{\eta,m}=\eta_{m}$ with $m=1,2,3$. In addition, $\left\langle \ldots\right\rangle _{0}$
is the expectation value with respect to the ground-state of $\mathcal{H}_{SQD}+\mathcal{H}_{DQD}$
at zero temperature. In the sequel, the term interacting/non-interacting
is used to account for the interaction between SQD and DQD and not
for the on-site interactions in the SQD. Also, we will present non-interacting
and interacting Green's functions by $g$ and $\mathcal{G}$, respectively.
Furthermore, because our Hamiltonian does not explicitly depend on
time, the Green's functions become functions of time differences only
and it is therefore more preferable to express them in the frequency
space by Fourier transformation.

The inclusion of interactions is performed by using the Dyson equation
through which the exact retarded Green's function of the system could
be determined by 
\begin{alignat}{1}
\mathcal{G}^{R}\left(\omega\right) & =\left[\mathcal{G}^{A}\left(\omega\right)\right]^{\dagger}\nonumber \\
 & =g^{R}\left(\omega\right)+g^{R}\left(\omega\right)\Sigma^{R}\left(\omega\right)\mathcal{G}^{R}\left(\omega\right),\label{eq:dyson_gr}
\end{alignat}
while the exact lesser Green's function has the form 
\begin{equation}
\mathcal{G}^{<}\left(\omega\right)=\mathcal{G}^{R}\left(\omega\right)\Sigma^{<}\left(\omega\right)\mathcal{G}^{A}\left(\omega\right),\label{eq:dyson_gl}
\end{equation}
where $\Sigma^{R,<}\left(\omega\right)$ stands for the total proper
retarded and lesser self-energies of the system. The greater Green's
function is then obtained using $\mathcal{G}^{>}\left(\omega\right)=\mathcal{G}^{A}\left(\omega\right)-\mathcal{G}^{R}\left(\omega\right)+\mathcal{G}^{<}\left(\omega\right)$.

\subsubsection{Green's functions of SQD}

In order to maximize the sensitivity of the SET, the energy level
of SQD should be tuned to the flank of the Coulomb blockade peak.
In this regime, the co-tunneling processes between SQD and leads become
dominant and the conventional sequential tunneling approximations
ceased to be applicable for describing the state of the SQD. Therefore
we use the infinite-U Lacroix approximation, which is believed to
consider co-tunnelings in the coulomb-blockade regime, in order to
obtain the Green's functions of SQD, $g_{d}^{R}$, which will be used
later on as building blocks of the self-energies. By using Eq.($17$)
of Ref.\onlinecite{Roermund}, we obtain the Fourier transform of
the diagonal elements of the SQD's retarded Green's function matrix
as
\begin{gather}
g_{d,\nu\nu}^{R}\left(\omega\right)=\frac{1-\left\langle n_{d,\bar{\nu}}\right\rangle +P_{\nu}\left(\omega\right)}{\omega+i\delta-\tilde{\varepsilon}_{d}+i\left(\Gamma_{L}+\Gamma_{R}\right)-Q_{\nu}\left(\omega\right)},\label{eq:g0r}
\end{gather}
where $\delta$ is an infinitesimal positive constant and $\Gamma_{L,R}\equiv\pi|t_{L,R}|^{2}\rho_{0}$
is the broadening of the SQD's energy level due to its coupling to
the leads in the standard wide-band limit in which the density of
states of the leads, $\rho_{0}$, is assumed to be independent of
energy. Furthermore,\begin{subequations}\label{eq:pq} 
\begin{equation}
P_{\nu}\left(\omega\right)=\underset{\alpha=L,R}{\sum}\frac{\Gamma_{\alpha}}{\pi}\int d\omega_{1}\frac{g_{d,\nu\nu}^{A}\left(\omega_{1}\right)f_{\alpha}\left(\omega_{1}\right)}{\omega+i\delta-\omega_{1}}
\end{equation}
and
\begin{gather}
Q_{\nu}\left(\omega\right)=\underset{\alpha=L,R}{\sum}\frac{\Gamma_{\alpha}}{\pi}\int d\omega_{1}\frac{\left(1+i\Gamma g_{d,\nu\nu}^{A}\left(\omega_{1}\right)\right)f_{\alpha}\left(\omega_{1}\right)}{\omega+i\delta-\omega_{1}},
\end{gather}
\end{subequations}where $f_{L,R}\left(\omega\right)=\theta\left(\mu_{L,R}-\omega\right)$
and $\theta\left(...\right)$ is the standard Heaviside-theta function.
For the non-interacting lesser Green's function of SQD, $g_{d}^{<}\left(\omega\right)$,
we have 
\begin{equation}
g_{d}^{<}\left(\omega\right)=g_{d}^{R}\left(\omega\right)\Sigma_{d}^{(NG)<}\left(\omega\right)g_{d}^{A}\left(\omega\right),
\end{equation}
 where $\Sigma_{d}^{(NG)<}$ is the lesser self-energy calculated
using the NG ansatz\cite{NGNG} 
\begin{gather}
\Sigma_{d}^{(NG)<}\left(\omega\right)=\left(\left[g_{d}^{R}\left(\omega\right)\right]^{-1}-\left[g_{d}^{A}\left(\omega\right)\right]^{-1}\right)\underset{\alpha=L,R}{\sum}\frac{\Gamma_{\alpha}f_{\alpha}\left(\omega\right)}{\Gamma}.\label{eq:NG}
\end{gather}

Using the self-energies of SQD, which are given in Appendix \ref{app_a},
the interacting retarded and lesser Green's functions of SQD can be
obtained as
\begin{equation}
\mathcal{G}_{d}^{R}\left(\omega\right)=g_{d}^{R}\left(\omega\right)+g_{d}^{R}\left(\omega\right)\Sigma_{d}^{(2nd)R}\left(\omega\right)\mathcal{G}_{d}^{R}\left(\omega\right)\label{eq:gdR}
\end{equation}
and
\begin{equation}
\mathcal{G}_{d}^{<}\left(\omega\right)=\mathcal{G}_{d}^{R}\left(\omega\right)\left(\Sigma_{d}^{(NG)<}\left(\omega\right)+\Sigma_{d}^{(2nd)<}\left(\omega\right)\right)\mathcal{G}_{d}^{A}\left(\omega\right).\label{eq:gdL}
\end{equation}

\subsubsection{Green's functions of DQD}

Using the method of equations of motion, the Fourier transform of
the non-interacting retarded Green's functions of DQD, $g_{\eta}^{R}$,
can be computed from the following set of nine equations 
\begin{alignat}{1}
\left(\omega+i\delta\right)g_{\eta,mn}^{R}= & \delta_{mn}+\delta_{1m}\tilde{\omega}_{0}g_{\eta,2n}^{R}-\nonumber \\
 & \delta_{2m}\left(\tilde{\omega}_{0}g_{\eta,1n}^{R}+\Delta g_{\eta,3n}^{R}\right)+\delta_{3m}\Delta g_{\eta,2n}^{R},
\end{alignat}
where $m,n=1,2,3$ and $\delta_{mn}$ is the Kronecker delta. The
solution of the above equations in matrix form is

\begin{gather}
g_{\eta}^{R}\left(\omega\right)=\left(\begin{array}{ccc}
\omega+i\delta & -i\tilde{\omega}_{0} & 0\\
i\tilde{\omega}_{0} & \omega+i\delta & i\Delta\\
0 & -i\Delta & \omega+i\delta
\end{array}\right)^{-1}.\label{eq:getta}
\end{gather}
Accordingly, the non-interacting lesser Green's function of DQD is
\begin{equation}
g_{\eta}^{<}\left(\omega\right)=-2iIm\left[g_{\eta}^{R}\left(\omega\right)\right]f\left(\omega\right),\label{eq:getteles}
\end{equation}
where $f\left(\omega\right)=\theta\left(-\omega\right)$.

Now we can use the self-energies of DQD, which are derived in Appendix
\ref{app_a}, to calculate the interacting retarded and lesser Green's
functions of DQD
\begin{equation}
\mathcal{G}_{\eta}^{R}\left(\omega\right)=g_{\eta}^{R}\left(\omega\right)+g_{\eta}^{R}\left(\omega\right)\Sigma_{\eta}^{(2nd)R}\left(\omega\right)\mathcal{G}_{\eta}^{R}\left(\omega\right)\label{eq:geR}
\end{equation}
and
\begin{equation}
\mathcal{G}_{\eta}^{<}\left(\omega\right)=\mathcal{G}_{\eta}^{R}\left(\omega\right)\Sigma_{\eta}^{(2nd)<}\left(\omega\right)\mathcal{G}_{\eta}^{A}\left(\omega\right).\label{eq:geL}
\end{equation}

\subsection{\label{sub:Physical-quantities}Physical quantities}

By using the definition of lesser Green's function, Eq.(\ref{eq:G_les}),
the expectation values of $n_{d}$ and $\tau_{3}$ are
\begin{equation}
\left\langle n_{d}\right\rangle =-\frac{i}{2\pi}\int d\omega Tr\left[\mathcal{G}_{d}^{<}\left(\omega\right)\right]\label{eq:nd}
\end{equation}
and
\begin{equation}
\left\langle \tau_{3}\right\rangle =-2\int\frac{d\omega}{2\pi}\mathcal{G}_{\eta,12}^{<}\left(\omega\right).\label{eq:sz}
\end{equation}
Furthermore, the average electric current through SQD in the steady-state
could be calculated by
\begin{equation}
I=-\frac{e}{\hbar}\int\frac{d\omega}{2\pi}\Gamma_{L}Tr[Im[\mathcal{G}_{d}^{<}\left(\omega\right)+2\mathcal{G}_{d}^{R}\left(\omega\right)f_{L}\left(\omega\right)],\label{eq:current}
\end{equation}
from which we can obtain the differential conductance of the SQD through
$G=\frac{dI}{dV_{b}}$. 

For future reference, we also define the ``signal differential conductance''
of the SQD\cite{barthel2010,petta}, which is defined as the difference
of the SQD's conductance in the presence of DQD and in the absence
of it. It is represented by 
\begin{equation}
\delta G=G_{\lambda\neq0}-G_{\lambda=0}.\label{eq:diff_signal}
\end{equation}

\section{\label{sec:Numerical-results}Results and discussions}

Here we present our numerical results for zero and finite bias voltages.
We calculate self-consistently the self-energies and the Green's functions
of the system (see Appendix \ref{app_b} for a brief outline of our
self-consistent calculations method). The calculations are performed
at zero-temperature $T=0$ and $\Gamma=\Gamma_{L}+\Gamma_{R}$ is
taken as unit of energy. Furthermore, we take $\hbar=e=c=1$. The
finite bias is established by considering a symmetric bias voltage
between two metallic lead as $\mu_{L}=-\mu_{R}=\frac{V_{b}}{2}$.
In zero bias, we check our results by comparing them with NRG results
which are obtained by utilizing ``NRG ljubljana''\cite{zitko} package.
In all NRG calculations we set the logarithmic discretization parameter
to $\Lambda=2$ and kept up to $1000$ states for each iteration diagonalizations.

\subsection{Spectral densities and average occupation values}

\begin{figure}
\includegraphics[width=8.6cm]{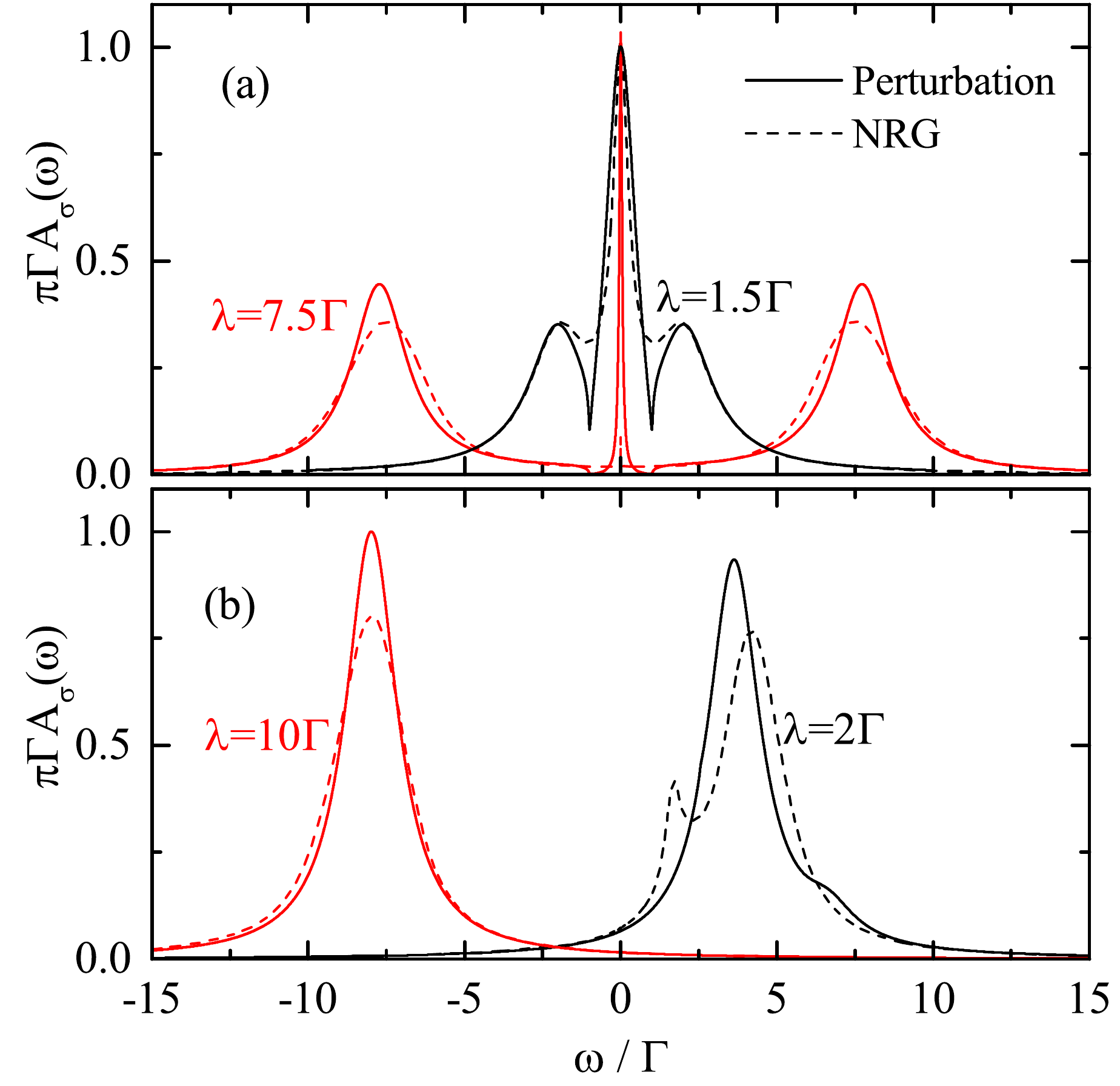}

\protect\caption{\label{figspec}Comparison of single spin spectral densities of SQD
calculated by perturbation method(full lines) with NRG method(dashed
lines) with $U=0$, $\Delta=\Gamma$ and $V_{b}=0$. For the cases
(a) $\varepsilon_{d}=0$, $2\lambda=\omega_{0}=3\Gamma,15\Gamma$
and (b) $\varepsilon_{d}=2\Gamma$, $\lambda=\frac{2}{3}\omega_{0}=2\Gamma,10\Gamma$.}
\end{figure}
In Fig.\ref{figspec}, we compare the single spin QD's local density
of states, $A_{\sigma}(\omega)=-\frac{1}{\pi}Im[\mathcal{G}_{d,\sigma\sigma}^{R}(\omega)]$,
obtained from our perturbative approach and NRG method. For simplicity,
we take $U=0$ and fix the value of $\Delta$ to $\Gamma$, while
we set different values to the $\lambda,\text{ }\omega_{0}\text{ and }\varepsilon_{d}$.
In Fig.\ref{figspec}(a), for the particle-hole symmetric case, $\varepsilon_{d}=0$
and $2\lambda=\omega_{0}$, we see good agreement between perturbative
results and NRG except the rate of narrowing of the central peak and
the height of the broad sidebands in the case of large $\lambda$s.
In Fig.\ref{figspec}(b), we show density of states for two particle-hole
asymmetric configurations. The position of the broad peaks are in
good agreement with NRG whereas their heights differ with it.

\begin{figure}
\includegraphics[width=8.6cm]{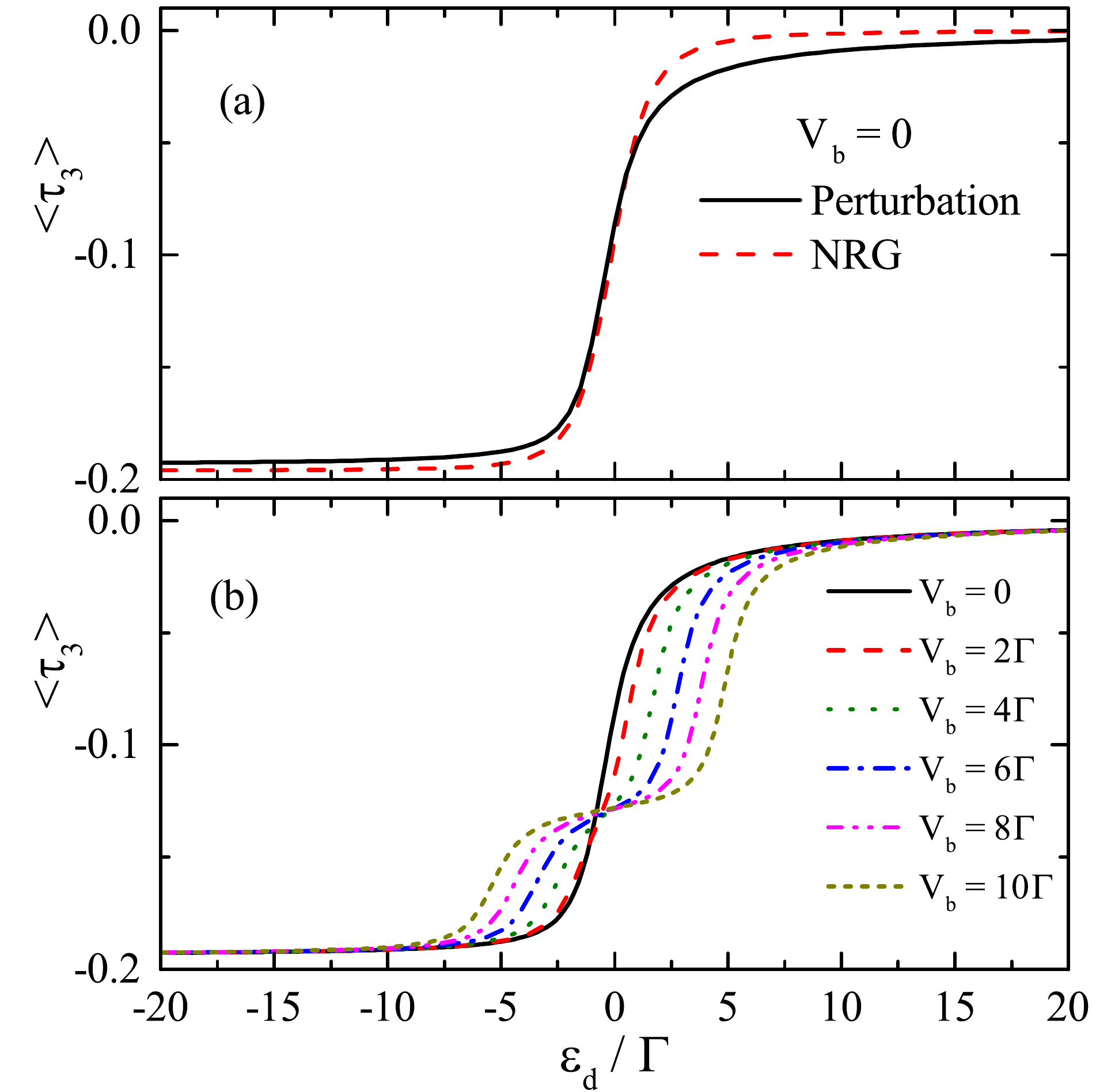}

\protect\caption{\label{figavg}Average values of $\tau_{3}$ with respect to $\varepsilon_{d}$
for $\omega_{0}=0$, $\Delta=10^{-1}\Gamma$ and $\lambda=10^{-2}\Gamma$
at (a) zero bias and (b) finite bias.}
\end{figure}
Next we consider the presence of large on-site interactions on the
SQD(infinite $U$) and focus on the weak-coupling parameter regime
where the condition $\lambda\ll\Delta\ll\Gamma$ is satisfied. We
set the energy difference between the two dots of DQD to zero, $\omega_{0}=0$,
and study the average occupation values of qubit for different gate
voltages of SQD in Fig.\ref{figavg}. Generally, it is expected that
the value of $\left\langle \tau_{3}\right\rangle $ becomes zero,
i.e. $\left(\left\langle n_{\pm}\right\rangle =\frac{1}{2}\right)$,
when there is no electron in SQD and by the presence of an electron
on SQD, the $\left\langle \tau_{3}\right\rangle $ acquires a negative
value to recover itself in the new potential energy of the qubit.
In Fig.\ref{figavg}(a) the average values $\left\langle \tau_{3}\right\rangle $,
obtained separately by our self-consistent method and NRG, are depicted
as a function of $\varepsilon_{d}$ for fixed $\Delta=10^{-1}$ and
$\lambda=10^{-2}$ when there is no applied bias. We see almost good
agreement with NRG. In the presence of finite bias voltages, as is
shown in Fig.\ref{figavg}(b), we see that by increasing bias voltages,
a step starts to appear in the average values $\left\langle \tau_{3}\right\rangle $
in the range $-\frac{V_{b}}{2}<\varepsilon_{d}<\frac{V_{b}}{2}$,
where the SQD has merely the same probability for being occupied or
unoccupied and therefore $\left\langle \tau_{3}\right\rangle $ acquires
a mid-value between zero and its minimum value.

\subsection{Differential conductance}

\begin{figure}
\includegraphics[width=8.6cm]{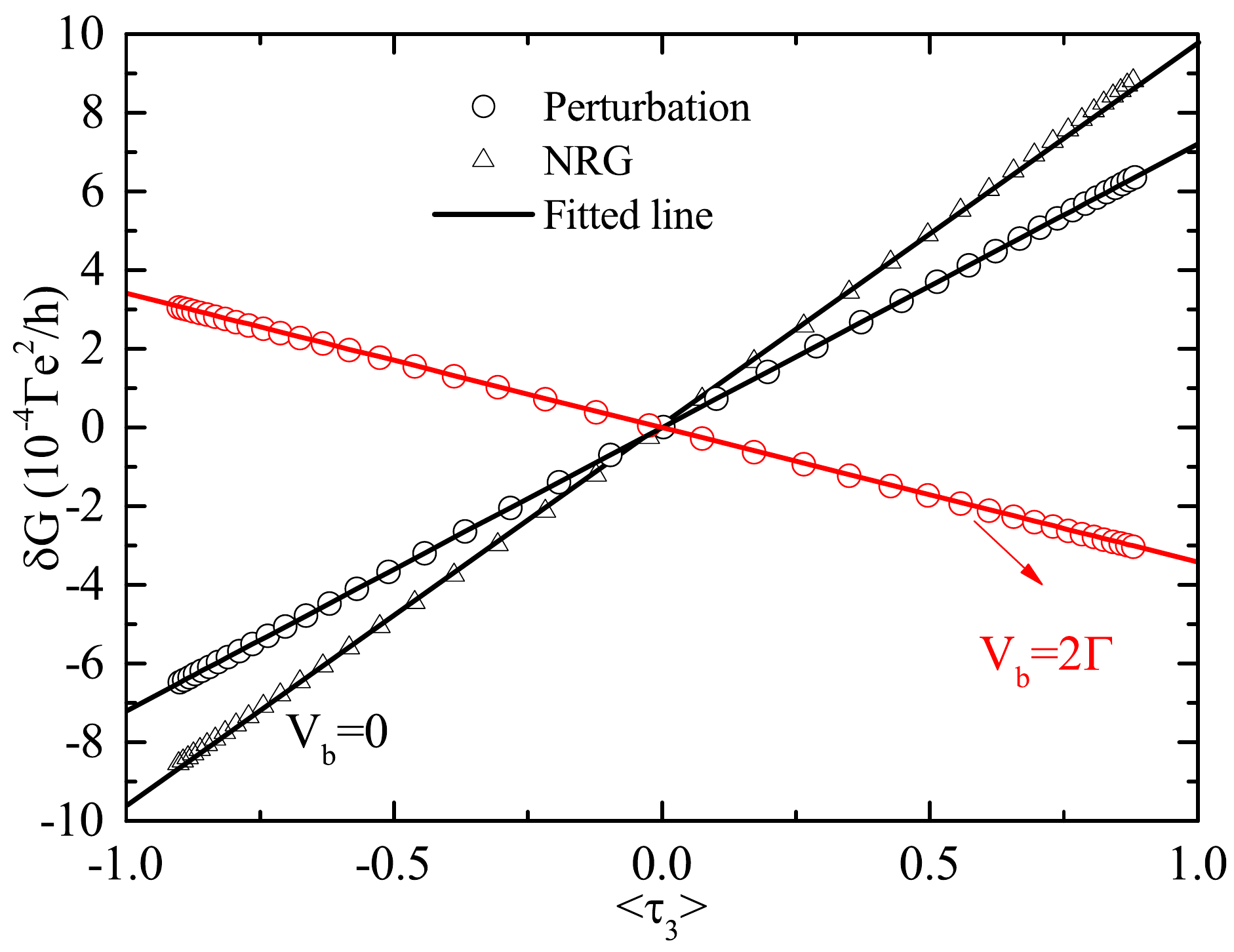}

\protect\caption{\label{figlin}Plot of $\delta G$ calculated by pertubative method(circles)
and NRG method(triangles) with respect to $\left\langle \tau_{3}\right\rangle $
in zero bias(black) and finite bias(red) voltages with $\varepsilon_{d}=0$,
$\Delta=10^{-1}\Gamma$, $\lambda=10^{-2}\Gamma$.}
\end{figure}
For a SQD with $U=0$ and in the weak-coupling parameter regime, we
can find an analytical expression for $\delta G$ which clearly shows
linear relation with $\left\langle \tau_{3}\right\rangle $ (see appendix
\ref{app_c} for a derivation). Our numerical results for this case
are also giving this linear relation perfectly (not shown here). On
the other hand, in the case of a SQD with infinite $U$, due to the
requirement of self-consistent calculations, obtaining an analytical
expression for $\delta G$ is seems to be impossible, at least in
the context of the NEGF formalism. Thus, in order to check the linear
dependence of $\delta G$ on $\left\langle \tau_{3}\right\rangle $,
we only concentrate on numerical results. In Fig.\ref{figlin}, our
numerical results for $\delta G$ as a function of $\left\langle \tau_{3}\right\rangle $
at zero and finite bias voltages are shown. We see that our perturbative
results(circles) are fitted entirely to a line which clearly demonstrates
the linear dependence of $\delta G$ on $\left\langle \tau_{3}\right\rangle $.
An important feature in Fig.\ref{figlin} is the linear dependence
of NRG results(triangles) for $\delta G$ with respect to $\left\langle \tau_{3}\right\rangle $
at zero bias. These NRG linear dependence could be thought of as a
complementary confirmation for our observations although its line
slope differs slightly from our perturbative results.

\begin{figure}
\includegraphics[width=8.6cm]{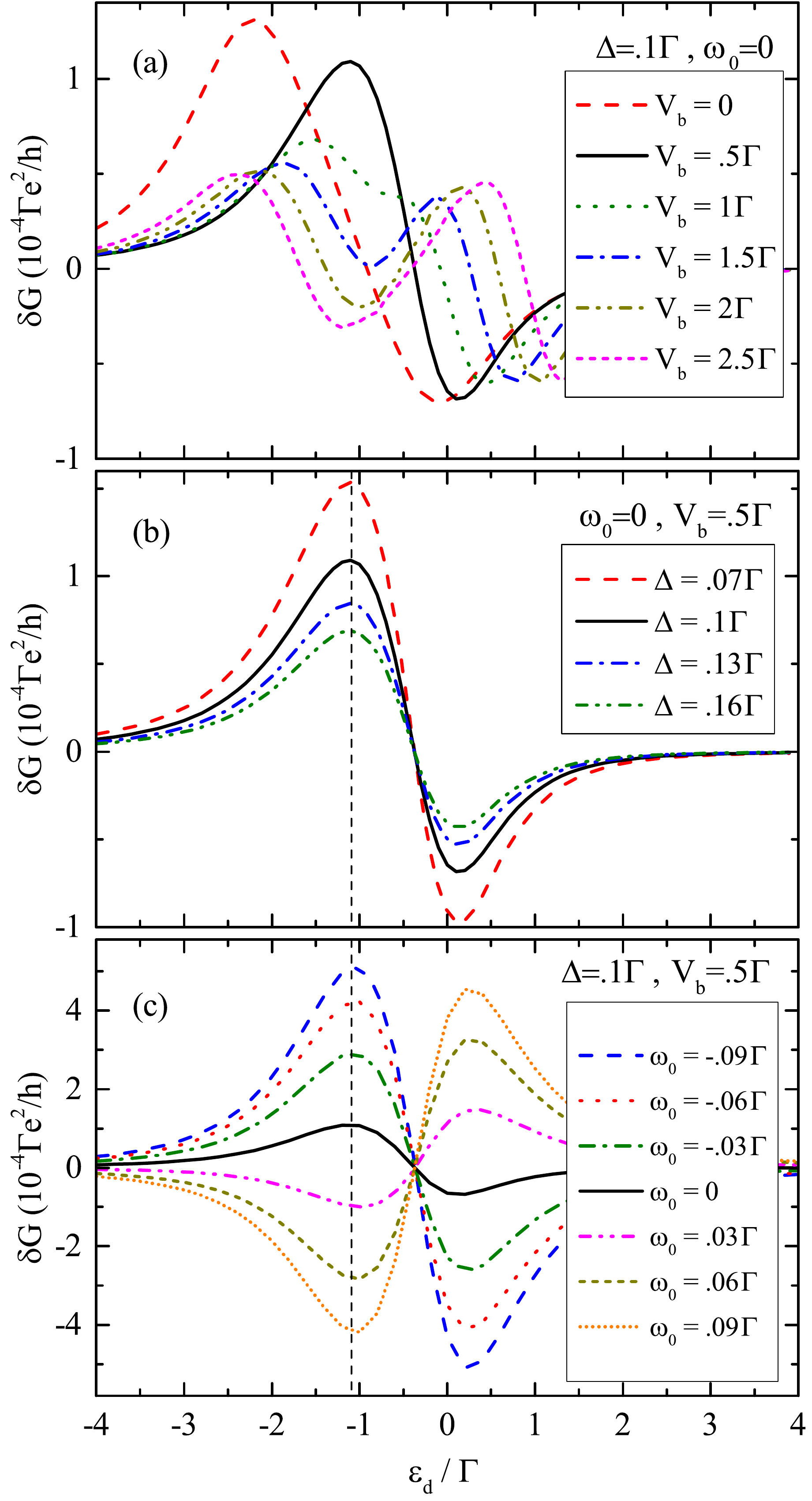}

\protect\caption{\label{figdif}Differential signal conductance of SQD as a function
of $\ensuremath{\varepsilon_{d}}$ for $\lambda=10^{-2}\Gamma$ with
respect to (a)$V_{b}$, (b)$\Delta$ and (c)$\omega_{0}$.}
\end{figure}
In Fig.\ref{figdif}, the dependence of $\delta G$ on the various
parameters of the system is shown. In Fig.\ref{figdif}(a), $\delta G$
is depicted as a function of $\varepsilon_{d}$ for different bias
voltages while the values of $\Delta$, $\lambda$ and $\omega_{0}$
are kept fixed. We see that the curves of $\delta G$ go from an infinitesimal
positive value for $\varepsilon_{d}\ll0$ to an infinitesimal negative
value for $\varepsilon_{d}\gg0$ while for intermediate values of
$\varepsilon_{d}$, they show some oscillations. By increasing the
value of $V_{b}$, the oscillations change from a ``one peak one
dip'' to a ``two peak two dip'' shape and also the positions of
the peaks/dips are pushed from the center. Another remarkable feature
is the decreasing of the amplitude of the $\delta G$ curves by increasing
the bias voltage. In other words, we predict that the amplitudes for
oscillations of signal differential conductances are maximized at
zero bias voltage. Next, we study the impact of changing $\Delta$
on $\delta G$ in Fig.\ref{figdif}(b). We see that increasing $\Delta$
has a reduction effect on $\delta G$ and decreases the amplitude
of the differential signal conductances. The other parameter of the
system which should have some effects on $\delta G$ is the energy
difference between the two quantum dots of the charge qubit ($\omega_{0}$).
In Fig.\ref{figdif}(c), we investigate the impact of different values
of $\omega_{0}$ on $\delta G$. We see that the position $\varepsilon_{d}$
of the peaks are almost intact, however, the amplitudes of the $\delta G$
curves change considerably by changing $\omega_{0}$.

\begin{figure}
\includegraphics[width=8.6cm]{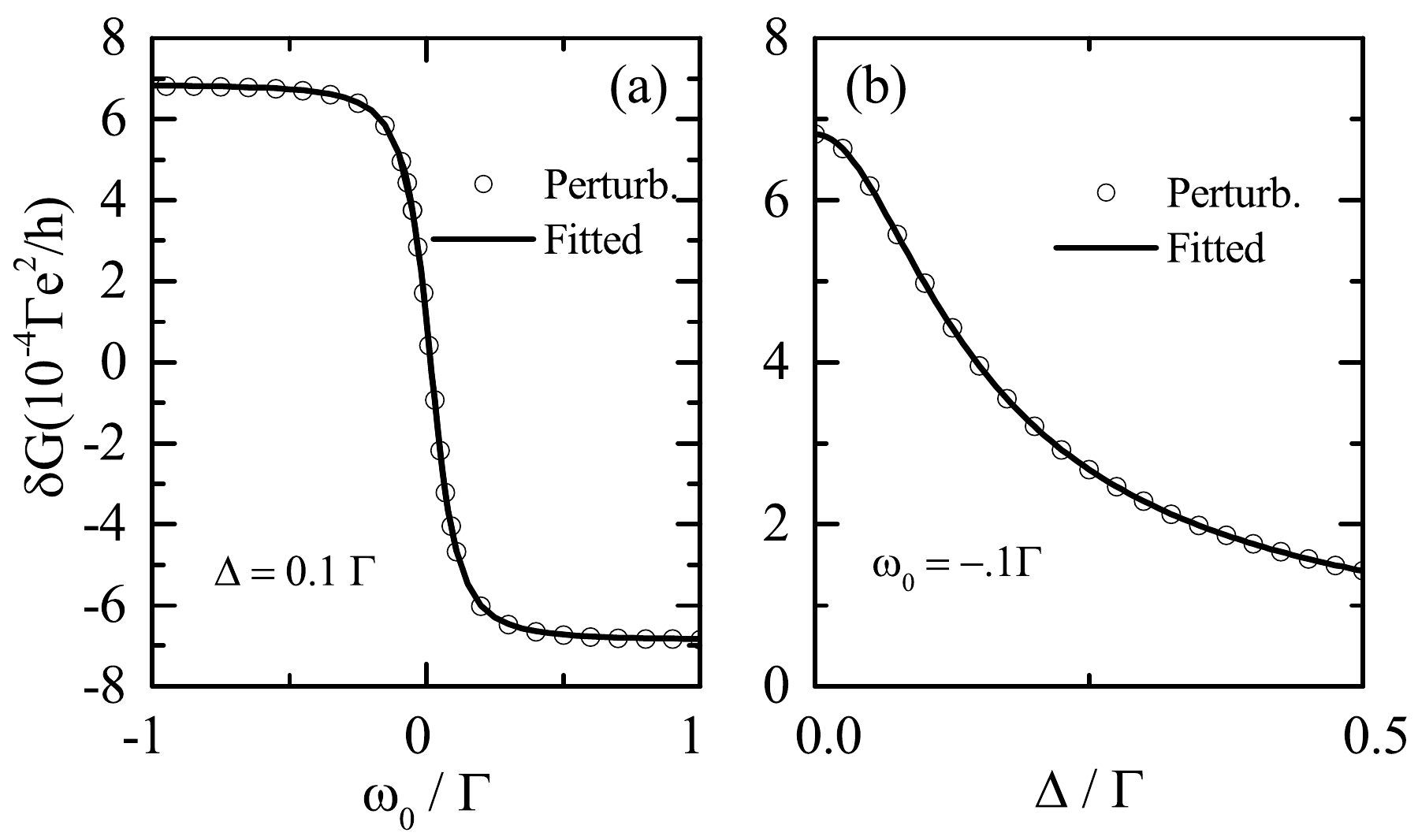}

\protect\caption{\label{figod}(a) Plot of $\delta G$ as a function of $\omega_{0}$
for $\Delta=.1\Gamma$, (b) Plot of $\delta G$ as a function of $\Delta$
for $\omega_{0}=-.1\Gamma$. Other parameters are $\ensuremath{\varepsilon_{d}}=-1.2\Gamma$,
$V_{b}=.5\Gamma$ and $\lambda=.01\Gamma$. Circles are perturbative
results and full lines are fittings to the function $f\left[\omega_{0},\Delta\right]$
with $a=1.6*10^{-2}$ and $c=-6.9*10^{-4}$.}
\end{figure}
In order to describe the aforementioned dependence of $\delta G$
on $\Delta$ and $\omega_{0}$, we focus on the peaks which are specified
in Figs.\ref{figdif}(b) and (c) by vertical dashed lines, and plot
the calculated values of $\delta G$ with respect to the $\omega_{0}$
and $\Delta$, respectively, in Figs.\ref{figod}(a) and (b). Surprisingly,
we see that both numerical data points in Figs.\ref{figod}(a) and
(b) are fitted perfectly to the function $f\left[\omega_{0},\Delta\right]=c\left(\omega_{0}-a\right)/\sqrt{\left(\omega_{0}-a\right)^{2}+\Delta^{2}}$.
We could intuitively interpret this behavior by the fact that the
ground-state expectation value of $\tau_{3}$ for an isolated charge
qubit, with the same configuration as in our model system, is equal
to $\left\langle \tau_{3}\right\rangle _{isolated}=-\omega_{0}/\sqrt{\omega_{0}^{2}+\Delta^{2}}$.
As a result the above functional form for $\delta G$ would be expected
to describe correctly the linear relation of $\delta G$ with the
ground-state expectation value of $\tau_{3}$ of a charge qubit which
is capacitively coupled to the SQD. 

From an experimental point of view, the above relation for $\delta G$
suggests a possible indirect measurement of the stationary-state value
of $\left\langle \tau_{3}\right\rangle $ by measuring $\delta G$.
By setting the value of $\omega_{0}$ to a very large value($\omega_{0}\rightarrow\pm\infty$),
the two end points of the lines in the Fig.\ref{figlin} are obtained.
Therefore one could find the value of constant $c$ by using the relation
$c=\left[\delta G\left(\omega_{0}\gg0\right)-\delta G\left(\omega_{0}\ll0\right)\right]/2$.
We emphasize the distinction between this measured value for $\left\langle \tau_{3}\right\rangle $
and the initial charge expectation value of the qubit (i.e. its value
before the measurement starts). Obviously, because all initial-state
informations are washed out by detector during readout process, the
stationary-state value of $\left\langle \tau_{3}\right\rangle $ is
by no means related to its initial value. Nevertheless, the study
of stationary-state properties of a qubit could be still desirable
in the sense that one can obtain certain informations about qubit-detector
coupled system and use them in performing manipulations or measurements
on the qubit\cite{makhlin2001quantum,averin2001continuous,korotkov2001output,ruskov2002quantum,ruskov2003spectrum,gurvitz2003relaxation,gilad2006qubit}.

At this point, it is interesting to compare our results for dependence
of $\delta G$ curves on $\Delta$ with the results of Ref.\onlinecite{hell2014}.
In that work, it is claimed that the overall shapes of $\delta G$
would not be altered to the first order in $\Delta$. To show that
our results are in accordance at some approximate level with the results
of Ref.\onlinecite{hell2014}, we could expand the function $\delta G=f\left[\omega_{0},\Delta\right]$
around small values of $\Delta$, then it is revealed that the correspondence
between $\delta G$ and $\Delta$ in Fig.\ref{figdif}(b) is actually
provided through next to the leading order in $\Delta$, i.e. $\delta G_{\Delta\rightarrow0}\approx c+\mathcal{O}\left(\Delta^{2}\right)$,
which is in accord with the above reference.

\section{\label{sec:conclusions}Conclusions}

We used the method of non-equilibrium Green's functions to study the
effect of electron-electron interaction between a SQD and a singly
occupied DQD(charge qubit) on their static and dynamic properties
at zero and finite bias voltages. To this end, we utilized the infinite-U
Lacroix approximation and the Majorana fermion representation of spin
operators to find the interacting Green's funcions of the system perturbatively
to the second order in the SQD-qubit coupling constant. We calculated
the Green's functions and self-energies of the system in a self-consistent
manner with which we could take into account the back-action effects
on the system. At zero bias, we checked the accuracy of our results
by comparing them with the NRG method. The agreement was good for
the density of states of SQD and the expectation value of difference
electron occupations of qubit($\left\langle \tau_{3}\right\rangle $).
We found a linear relation between the differential conductance of
SQD($\delta G$) and stationary-state expectation value of $\left\langle \tau_{3}\right\rangle $.
Concerning with this linear relation, we gave NRG results at zero
bias, as a support for our perturbative results, from which perfect
linear relation was observed. We also investigated the dependency
of $\delta G$ on various parameters of the system such as $V_{b}$,
$\Delta$, $\omega_{0}$ and $\varepsilon_{d}$. We found that the
$\delta G$ curves are best pronounced at zero bias voltage and their
amplitudes are decreased relatively by increasing bias voltages. Furthermore,
we found an approximate functional form for $\delta G$ with respect
to $\Delta$ and $\omega_{0}$. By using this analytical expression,
we became able to describe the reason why the authors of Ref.\onlinecite{hell2014}
stated that the $\delta G$ curves are not dependent on the values
of $\Delta$.
\begin{acknowledgments}
We acknowledge fruitful discussions with Farshad Ebrahimi. We also
thank Amir Eskandari Asl, Babak Zare Rameshti, Micheal Hell, Rok Zitko
and Pablo Cornaglia for their useful comments.
\end{acknowledgments}

\appendix

\section{\label{app_a}Expressions for self-energies of SQD and DQD}

In this appendix, we give the expressions for self-energies of SQD
and DQD, respectively, due to the interaction Hamiltonian $\mathcal{H}_{I}$.
The first order self-energies are identically zero for both SQD and
DQD because we have taken into account their effect in the non-interacting
Green's functions.

The SQD's second order self-energies are\begin{subequations}\label{eq:selfg}
\begin{alignat}{1}
\Sigma_{d}^{(2nd)R}\left(\omega\right) & =\lambda^{2}\int\frac{d\omega_{1}}{2\pi}[g_{d}^{<}\left(\omega\right)\Phi^{R}\left(\omega-\omega_{1}\right)\nonumber \\
 & +g_{d}^{R}\left(\omega\right)\Phi^{<}\left(\omega-\omega_{1}\right)+g_{d}^{R}\left(\omega\right)\Phi^{R}\left(\omega-\omega_{1}\right)]
\end{alignat}
and
\begin{gather}
\Sigma_{d}^{(2nd)<}\left(\omega\right)=\lambda^{2}\int\frac{d\omega_{1}}{2\pi}g_{d}^{<}\left(\omega\right)\Phi^{<}\left(\omega-\omega_{1}\right),
\end{gather}
\end{subequations}where\begin{subequations}
\begin{align}
\Phi^{R}(\omega)=\int\frac{d\omega_{1}}{2\pi}[ & g_{\eta,11}^{<}\left(\omega+\omega_{1}\right)g_{\eta,22}^{A}\left(\omega_{1}\right)+\nonumber \\
 & g_{\eta,11}^{R}\left(\omega+\omega_{1}\right)g_{\eta,22}^{<}\left(\omega_{1}\right)-\nonumber \\
 & g_{\eta,12}^{<}\left(\omega+\omega_{1}\right)g_{\eta,21}^{A}\left(\omega_{1}\right)-\nonumber \\
 & g_{\eta,12}^{R}\left(\omega+\omega_{1}\right)g_{\eta,21}^{<}\left(\omega_{1}\right)]
\end{align}
and
\begin{alignat}{1}
\Phi^{<}(\omega)=\int\frac{d\omega_{1}}{2\pi}[ & g_{\eta,11}^{<}\left(\omega+\omega_{1}\right)g_{\eta,22}^{>}\left(\omega_{1}\right)-\nonumber \\
 & g_{\eta,12}^{<}\left(\omega+\omega_{1}\right)g_{\eta,21}^{>}\left(\omega_{1}\right)].
\end{alignat}
\end{subequations}

For DQD, the second order self-energies are
\begin{align}
\Sigma_{\eta}^{(2nd)R,<}\left(\omega\right) & =\left(\begin{array}{ccc}
F_{22}^{R,<}\left(\omega\right) & F_{21}^{R,<}\left(\omega\right) & 0\\
F_{12}^{R,<}\left(\omega\right) & F_{11}^{R,<}\left(\omega\right) & 0\\
0 & 0 & 0
\end{array}\right),\label{eq:selfeta}
\end{align}
where\begin{subequations}\label{eerr} 
\begin{alignat}{1}
F_{mn}^{R}\left(\omega\right)=\lambda^{2}\underset{\nu=\uparrow,\downarrow}{\sum}\int\frac{d\omega_{1}}{2\pi}[ & g_{\eta,mn}^{<}\left(\omega\right)\Pi_{\nu}^{R}(\omega-\omega_{1})+\nonumber \\
 & g_{\eta,mn}^{R}\left(\omega\right)\Pi_{\nu}^{<}(\omega-\omega_{1})+\nonumber \\
 & g_{\eta,mn}^{R}\left(\omega\right)\Pi_{\nu}^{R}(\omega-\omega_{1})]
\end{alignat}
and
\begin{equation}
F_{mn}^{<}\left(\omega\right)=\lambda^{2}\underset{\nu=\uparrow,\downarrow}{\sum}\int\frac{d\omega_{1}}{2\pi}g_{\eta,mn}^{<}\left(\omega\right)\Pi_{\nu}^{<}(\omega-\omega_{1}).
\end{equation}
\end{subequations}In Eqs.(\ref{eerr}), the functions $\Pi_{\nu}^{R,<}\left(\omega\right)$
are given by\begin{subequations}
\begin{alignat}{1}
\Pi_{\nu}^{R}(\omega)=\int\frac{d\omega_{1}}{2\pi}[ & g_{d,\nu\nu}^{<}\left(\omega+\omega_{1}\right)g_{d,\nu\nu}^{A}\left(\omega_{1}\right)+\nonumber \\
 & g_{d,\nu\nu}^{R}\left(\omega+\omega_{1}\right)g_{d,\nu\nu}^{<}\left(\omega_{1}\right)]
\end{alignat}
and
\begin{equation}
\Pi_{\nu}^{<}(\omega)=\int\frac{d\omega_{1}}{2\pi}g_{d,\nu\nu}^{<}\left(\omega+\omega_{1}\right)g_{d,\nu\nu}^{>}\left(\omega_{1}\right).
\end{equation}
\end{subequations}

\section{\label{app_b}Self-consistent calculations}

In our numerical results, we have calculated the set of four unknown
quantities ($P_{\nu}\left(\omega\right)$, $Q_{\nu}\left(\omega\right)$,
$\left\langle n_{d,\nu}\right\rangle $ and $\left\langle \tau_{3}\right\rangle $)
by solving self-consistently the Eqs.(\ref{eq:pq}), (\ref{eq:nd})
and (\ref{eq:sz}). This way we assure that the back-action effects
are correctly taken into account in the results. We used the following
scheme:

($\mathit{i}$) We start with an initial guess for $\left\langle n_{d}\right\rangle $
and $\left\langle \tau_{3}\right\rangle $ and set $P\left(\omega\right)=Q\left(\omega\right)=0$
and then compute $g_{d}^{R}\left(\omega\right)$ from Eq.(\ref{eq:g0r}).

($\mathit{i}i$) We calculate $P\left(\omega\right)\text{ and }Q\left(\omega\right)$
from computed $g_{d}^{R}\left(\omega\right)$ and use them to obtain
a new $g_{d}^{R}\left(\omega\right)$. We iterate this step until
convergence over $g_{d}^{R}\left(\omega\right)$ is attained.

($\mathit{i}ii$) Using the calculated $g_{d}^{R,<}\left(\omega\right)$
and $g_{\eta}^{R,<}\left(\omega\right)$, the self-energies are calculated
straightforwardly and then we use the interacting lesser Green's functions,
Eqs.(\ref{eq:gdL}) and (\ref{eq:geL}), to calculate new $\left\langle n_{d}\right\rangle $
and $\left\langle \tau_{3}\right\rangle $.

We iterate these three steps until convergence over $\left\langle n_{d}\right\rangle $
and $\left\langle \tau_{3}\right\rangle $ is attained.

\section{\label{app_c}Analytical expression for $\delta G$}

For a SQD with $U=0$, it is possible to obtain an analytical expression
for the linear relation between $\delta G$ and $\left\langle \tau_{3}\right\rangle $.
To this end, we expand Eq.(\ref{eq:diff_signal}) to the first order
in $\lambda$, i.e. $\delta G_{\lambda\rightarrow0}\approx\lambda\left.\frac{\partial G_{\lambda\neq0}}{\partial\lambda}\right|_{\lambda=0}$,
and then using Eq.(\ref{eq:current}) and $\mathcal{G}_{d}^{R}\left(\omega\right)=\left[\omega-\varepsilon_{d}-\lambda\left\langle \tau_{3}\right\rangle +i\Gamma-\Sigma_{d}^{(2nd)R}\left(\omega\right)\right]^{-1}$,
we obtain 
\[
\delta G_{\lambda\rightarrow0}\approx\lambda\underset{V=\pm\frac{V_{b}}{2}}{\sum}\left(\frac{\Gamma(\varepsilon_{d}+V)}{\pi\left(\left(\varepsilon_{d}+V\right)^{2}+\Gamma^{2}\right)^{2}}\right)\left\langle \tau_{3}\right\rangle +\mathcal{O}\left(\lambda^{2}\right).
\]
One immediate consequence of this expression is that even at the zero
bias voltage there would be an obvious conductance difference in the
system, i.e. $\delta G_{V_{b}=0}\approx\lambda\frac{2\Gamma\varepsilon_{d}}{\pi\left(\varepsilon_{d}^{2}+\Gamma^{2}\right)^{2}}\left\langle \tau_{3}\right\rangle +\mathcal{O}\left(\lambda^{2}\right)$. 

\bibliographystyle{unsrt}
\bibliography{6E__manuscripts_Perturbative_approach_to_the_capacitive_interaction_between_a__sensor_refs_new}

\end{document}